\documentclass[aps,prl,twocolumn,showpacs]{revtex4-1}

\usepackage{amsfonts,amssymb,amsmath}

\usepackage{times}
\usepackage{graphicx}
\usepackage{bm}
\usepackage{color}
\usepackage{hyperref}

\usepackage{xfrac}
\usepackage{ulem}
\usepackage{soul}
\usepackage{epstopdf}

\newcommand{\bee}{\begin{equation}}
\newcommand{\eee}{\end{equation}}
\newcommand{\ba}{\begin{eqnarray}}
\newcommand{\ea}{\end{eqnarray}}

\def\ARED#1{{\textcolor{red}{#1}}}        

\begin{document}
\title{\bf  An exact result in strong wave turbulence of thin elastic plates } 
\author{Gustavo D\"uring${}^{1}$ and Giorgio Krstulovic${}^{2}$}

\affiliation {${}^1$ Facultad de F\'isica, Pontificia Universidad Cat\'olica de Chile, Casilla 306, Santiago, Chile\\
 ${}^2$ Universit\'e de la C\^ote d'Azur, OCA, CNRS, Lagrange, France. B.P. 4229, 06304 Nice Cedex 4, France\\
}
\date{\today}

\begin{abstract}
 An exact result concerning the energy transfers between non-linear waves of thin elastic plate is derived. Following Kolmogorov's original ideas in hydrodynamical turbulence, but applied to the F\"oppl-von K\'arm\'an equation for thin plates, the corresponding K\'arm\'an-Howarth-Monin relation and an equivalent of the $\sfrac{4}{5}$-Kolmogorov's law is derived. A third-order structure function involving increments of the amplitude, velocity and the Airy stress function of a plate, is proven to be equal to $-\varepsilon\, \ell$, where $\ell$ is a length scale in the inertial range at which the increments are evaluated and $\varepsilon$ the energy dissipation rate. Numerical data confirm this law. In addition, a useful definition of the energy fluxes in Fourier space is introduced and proven numerically to be flat in the inertial range. The exact results derived in this Letter are valid for both, weak and strong wave-turbulence. They could be used as a theoretical benchmark of new wave-turbulence theories and to develop further analogies with hydrodynamical turbulence.
 \end{abstract}

\maketitle

Hydrodynamic turbulence (HDT) is considered as a prototype of systems far from equilibrium. The understanding of its statistical properties has challenged over the last century physicists and mathematicians. Today, few exact results are available. The main difficulty is the strong nonlinearity and the lack of a small parameter. The phenomenological description of turbulence is based on the idea proposed by Richardson, in which energy is transferred along scales at a constant flux \cite{frisch1996turbulence}. This process is seen as a cascade of \textit{eddies} that starts at large scales, where energy is injected, and ends at small scales, where it is dissipated. The seminal works of Kolmogorov are the most general results we have nowadays. In particular, its celebrated $\sfrac{4}{5}$-law \cite{kolmogorov1941FourFith}, which gives an explicit expression for the third order moment of the velocity increments, provides a benchmark for any theoretical description of turbulence. This exact result has been generalised to other transport-like systems such as a passive scalar transported by a incompressible turbulent flow \cite{yaglom1949local}, magnetohydrodynamic turbulence \cite{PouquetPolitanoPRE1998} and rotating turbulence \cite{GaltierRotatingTrubulence}, among others. Exact results are rare in turbulence, what makes Kolmogorov $\sfrac{4}{5}$-\textit{law} one of the most important prediction in HDT.

During the sixties an important theoretical breakthrough occurred with the development of the theory of (weak) wave-turbulence \cite{ZakharovBook}. Due to non-linear interactions, waves transfer energy along scales like in a cascade process. In analogy with HDT, this out-of-equilibrium phenomenon was named wave-turbulence (WT). In contrast with HDT, for weak WT exists a small parameter which allows for a natural perturbation expansion \cite{benney,benneynewell69,RevGeo}. 
The statistical properties of weakly nonlinear wave systems have been thus proven to evolve through a kinetic equation for the second order moments of the wave amplitudes \cite{hasselmann}. Many different systems such as waves in plasma \cite{Zakplasma67,Sagdeev,kuznetso,galtier2000}, spin waves in solids \cite{spinwave1,Lvovbook}, surface waves in fluids \cite{hasselmann,zakgrav66,zakcap67,benney,benneynewell69} and nonlinear optics \cite{Dyachenko-92, DuringBEC} among others, have been shown to follow similar kinetic equations in the weakly nonlinear regime. Moreover, Zakharov has shown that  stationary, out-of-equilibrium power-law solutions, naturally emerge from the kinetic equation \cite{Zakplasma67}. Such solutions are related to the flux of conserved quantities, similarly to Kolmogorov prediction for the kinetic energy spectrum in HDT. In the last decade the interest in WT has been boosted by the development of new experimental settings \cite{FauveG, Mordant, Mordant2, Arezki, 4wavecapi,Fauve1,CFalcon09,Denissenko07,Bortolozzo09} and new numerical simulations \cite{placa2006,Deike14,CMMT2,Yokoyama13} that have been able to test WT predictions.   Particularly fruitful has been the development of WT for thin elastic plates  \cite{placa2006}. From both sides, numerical and experimental, thin elastic plates has shown to be one of the ideal settings to address fundamental issues of the theory of WT and its breakdown \cite{Mordant,Mordant2,Arezki,Cadot2008,Touze2012,humbert2013wave,Miquel2013,Auliel2015,Yokoyama13,during2016}  (for a review see \cite{during2017wave,cadot2016book}). 

Until recently, HDT has been considered a rather different problem to the one of WT. However, in the last years the observation of an intermittent behaviour in WT experiments on gravity-capillary waves \cite{Fauve1} and in simulations of elastic plates \cite{ChibbaroJosserandPlateIntermittency}, has suggested that a closer connection with HDT  could exist when the non-linearity of waves is strong enough \cite{Newell}.  {Unfortunately, results are very scarce in this regime \cite{FalkoNLS,Connaughton2007PRL}. }
What are the concepts and theoretical tools that can be borrowed from HDT to be applied in WT, or vice versa, remains an open question.

In this Letter, we provide a bridge between strong and weak WT in elastic plates deriving an exact result concerning the energy transfers.  We derive the corresponding K\'arm\'an-Howarth-Monin relation and an exact result for a third order structure function that is equivalent to the $\sfrac{4}{5}$-Kolmogorov's law for HDT. We call this result, as it will be naturally motivated  later, the $1$-\textit{law} of thin elastic plates. Remarkably, unlike other systems where a K\'arm\'an-Howarth-Monin relation has been derived, thin elastic plates dynamics is not given by a transport equation. We then provide numerical data corroborating the $1$-\textit{law} of thin elastic plates. The results presented in this Letter are valid independently of the strength of the nonlinear interaction of waves, and reduce one step further the gap between HDT and elastic WT phenomena.
 
To model the vibration of an elastic plate, we use the dynamical version of the F\"oppl-von K\'arm\'an (FvK) equations for the vertical amplitude of the deformation $\zeta(x,y,t)$ and the Airy stress function $\chi(x,y,t):$
\begin{eqnarray} 
\rho\frac{\partial^2 \zeta}{\partial t^2} &=& - \frac{l^2 E}{4}\Delta^2\zeta +
\{\zeta,\chi\}  +\mathcal{F}-\nu(-\Delta)^{\frac{\alpha}{2}}\frac{\partial \zeta}{\partial t}
\label{foppl0}\\
\Delta^2\chi &=&- \frac{E}{2}\{\zeta,\zeta\},
\label{foppl1}
\end{eqnarray}
where $\l=\frac{h}{\sqrt{3(1-\sigma^2)}}$, with $h$ the thickness of the elastic sheet and $\sigma$ the Poisson
ratio.  The material has a  mass density$\rho$, a Young modulus $E$ and a damping coefficient $\nu$.
$\Delta$ is the usual Laplacian and the bracket $\{\cdot,\cdot\}$ is defined by
$\{f,h\}\equiv f_{xx}h_{yy}+f_{yy}h_{xx}-2f_{xy}h_{xy}$. A fundamental property to derive the $1$-\textit{law}, as we will see below, is that the bracket can be written as a total divergence
\bee
\label{div}
\{f,h\}=-\nabla\cdot {\bm J}_{[ f ,h]}=-\nabla\cdot {\bm J}_{[ h ,f]},
\eee 
where{
\begin{equation} 
{\bm J}_{[ f(x,y) ,h(x,y)]}=\begin{pmatrix} f_y h_{yx}-f_{x}h_{yy} \\ f_x h_{xy} - f_{y} h_{xx}\end{pmatrix}.
\end{equation} }
The last two terms in \eqref{foppl0} are the external forcing $\mathcal{F}$ and the small-scale ($\alpha>0)$ dissipation respectively.

Equation (\ref{foppl1}) for  the Airy stress function
$\chi(x,y,t)$ may be seen as the compatibility equation for the in--plane stress tensor which follows the dynamics.  When $\mathcal{F}$ and $\nu$ vanish, the FvK equations are conservative and derive from the Hamiltonian
\bee
\label{Hamiltonian0}
 {H} = h \int   \frac{\rho}{2} \dot \zeta^2+\frac{l^2E}{8}(\Delta\zeta)^2  -  \frac{1}{2E}(\Delta\chi)^2 - \frac{1}{2} \chi \{\zeta,\zeta\} d{\bm r}. 
 \eee
 Integrating by parts the last term in (\ref{Hamiltonian0}) and using (\ref{foppl1}), the Hamiltonian can be rewritten as $H=h \int {\mathcal{H}}({\bm r})d{\bm r}$ where the energy density $E({\bm r})$ is defined as
 \bee
 \label{Energy}
\ARED{\mathcal{H}}({\bm r})= \frac{\rho}{2} \dot \zeta^2+\frac{l^2E}{8}(\Delta\zeta)^2+ \frac{1}{2E}(\Delta\chi)^2.
 \eee
 The first term in (\ref{Energy}) corresponds to the kinetic energy, whereas the other two have a purely geometric origin. The middle term is the bending energy which is related to mean curvature and the last one is the nonlinear stretching coming from the Gaussian curvature.

We consider in the following an elastic plate in a turbulent state driven by the external forcing at large scales and energy dissipated at small scales by some damping mechanisms \cite{humbert2013wave}.

We turn now to the derivation of the K\'arm\'an-Howarth-Monin relation for statistically homogenous elastic plates.  As usual \cite{frisch1996turbulence}, we shall introduce the correlation functions
\ba
\label{kineticE}
\mathcal{E}_\text{kin}({\bm \ell})&=&  \frac{\rho}{2}\langle  \dot \zeta({\bm r}) \dot \zeta({\bm r}')\rangle,\\
\label{bendingE}
\mathcal{E}_\text{ben}({\bm \ell})&=&   \frac{l^2E}{8}\langle\Delta_{\bm r}\zeta({\bm r})\Delta_{{\bm r}'}\zeta({\bm r}')\rangle,\\
\label{stretching}
\mathcal{E}_\text{stret}({\bm \ell})&=& \frac{1}{2E}\langle\Delta_{\bm r}\chi({\bm r})\Delta_{{\bm r}'}\chi({\bm r}')\rangle,
\ea
where $\Delta_{\bm r}$ represent the Laplacian with respect to  ${\bm r}$  and ${\bm \ell}~=~{\bm r}'-{\bm r}$. The brackets  $\langle\,\,\rangle$ stand for ensemble average. Statistical homogeneity guarantees that two-point correlation functions depend only on the distance ${\bm \ell}$. Notice that taking the limit $\ell\rightarrow 0$ the correlation function (\ref{kineticE}), (\ref{bendingE}) and (\ref{stretching}) correspond to the mean kinetic, bending and stretching energy respectively defined in (\ref{Energy}).

To establish a relation between the energy flux and the statistical properties of the plate we need to take the time derivatives of  (\ref{kineticE}), (\ref{bendingE}) and (\ref{stretching}).  
The simplest term is obtained from \eqref{bendingE} after a direct calculation: 
\bee
\label{dotBending}
\mathcal{\dot{E}}_\text{ben}({\bm \ell})=\frac{l^2E}{8} \frac{d}{d t} \left(\Delta^2_{\bm \ell} \,\langle \zeta \zeta' \rangle\right)
\eee
where $\zeta'=\zeta({\bm r}')$ and $\zeta=\zeta({\bm r})$. To derive (\ref{dotBending}) we have used the property that for statistically homogenous systems, an arbitrary function ${\bf g}( {\bm r},{\bm r}')$  satisfies the following relation
\begin{equation}
\langle \nabla_{{\bm r}'}{\bf g}( {\bm r},{\bm r}')\rangle=-\langle \nabla_{\bm r}{\bf g}( {\bm r},{\bm r}')\rangle=\nabla_{\bm \ell}\langle {\bf g}( {\bm r},{\bm r}')\rangle.
 \end{equation}

To calculate the time derivative of (\ref{kineticE}) we make use of the equations of motions (\ref{foppl0}). A straightforward calculation using the definition (\ref{div}) leads to 
\ba
\label{dotKinetic}
\mathcal{\dot{E}}_\text{kin}({\bm \ell})&=&\frac{1}{2}\nabla_{\bm \ell}\cdot \left(\langle {\bm J}_{[\chi ,\zeta]}\dot{\zeta}'\rangle-\langle {\bm J}_{[\chi' ,\zeta']}\dot{\zeta}\rangle \right)-\mathcal{\dot{E}}_\text{ben}({\bm \ell})\nonumber\\
&+&\frac{1}{2}\langle \dot{\zeta} \mathcal{F}'+\dot{\zeta}' \mathcal{F}\rangle-\nu (-\Delta_{\bm \ell})^{\frac{\alpha}{2}} \langle \dot{\zeta} \dot{\zeta}' \rangle.
\ea
{The flux of stretching energy \eqref{stretching} requires some algebra. Using Eq.(\ref{foppl1}) and the identity $\langle \{f,g\}h\rangle=\langle  \{h,f\}g \rangle$ it gives 
\ba
\label{dotStretching}
\mathcal{\dot{E}}_\text{stret}({\bm \ell})&=&\frac{1}{2E}\left(\langle \chi \frac{d}{dt}\Delta^2\chi'\rangle+\langle\chi' \frac{d}{dt}\Delta^2\chi \rangle\right)\nonumber\\
&=&-\frac{1}{2}\left(\langle \chi \{\zeta',\dot{\zeta'}\}\rangle+\langle\chi'  \{\zeta,\dot{\zeta}\} \rangle\right)\nonumber\\
&=&\frac{1}{2}\nabla_{\bm \ell}\cdot\left(\langle {\bm J}_{[\chi',\zeta]}\dot{\zeta}\rangle-\langle {\bm J}_{[\chi ,\zeta']}.\dot{\zeta}'\rangle \right)
\ea 
} 
The next step to obtain a K\'arm\'an-Howarth-Monin relation, is to introduce the increment of a field. For an arbitrary function $g({\bm r})$ its increment is defined as $\delta g=g({\bm r}')-g({\bm r})$. We shall notice the following identity
\ba
\label{SFunction}
\langle  {\bm J}_{[\delta\chi ,\delta\zeta]}\delta\dot{\zeta}\rangle\nonumber=&&\langle {\bm J}_{[\chi, \zeta]}\dot{\zeta}'\rangle-\langle {\bm J}_{[\chi' ,\zeta']}\dot{\zeta}\rangle
+\langle {\bm J}_{[\chi' ,\zeta]}\dot{\zeta}\rangle-\langle {\bm J}_{[\chi,\zeta']}\dot{\zeta}'\rangle\\
&&+\,\langle {\bm J}_{[\chi,\zeta']}\dot{\zeta}\rangle-\langle {\bm J}_{[\chi' ,\zeta]}\dot{\zeta}'\rangle.
\ea
One can easily show that the divergence of the last two terms in the latter expression vanish identically. Therefore,
collecting the expression obtained in (\ref{dotBending}), (\ref{dotKinetic}), (\ref{dotStretching}) and using (\ref{SFunction}), we finally find the K\'arm\'an-Howarth-Monin relation for statistically homogenous WT in thin elastic plates 
\bee
\label{KHM}
\frac{1}{2}\nabla_{\bm \ell}\cdot  \langle  {\bm J}_{[\delta\chi,\delta\zeta]}\delta\dot{\zeta}\rangle=\mathcal{\dot{E}}(\ell)-\frac{1}{2}\langle \dot{\zeta} \mathcal{F}'+\dot{\zeta}' \mathcal{F}\rangle+\gamma (-\Delta_{\bm \ell})^{\frac{\alpha}{2}} \langle \dot{\zeta} \dot{\zeta}' \rangle
\eee
where $\mathcal{\dot{E}}({\bm \ell})=\mathcal{\dot{E}}_\text{kin}({\bm \ell})+\mathcal{\dot{E}}_\text{ben}({\bm \ell})+\mathcal{\dot{E}}_\text{stret}({\bm \ell})$.  In a statistically stationary turbulent state, if the injection and dissipation scales are well separated, an inertial range exist. Inside this inertial range the right-hand side of equation (\ref{KHM}) becomes minus the energy flux $\varepsilon$, which is assumed to be finite and constant as in HDT \cite{frisch1996turbulence}. Therefore the K\'arm\'an-Howarth-Monin relation  (\ref{KHM}) reduces to
\begin{equation}
\frac{1}{2}\nabla_{\bm \ell}\cdot  \langle  {\bm J}_{[\delta\chi,\delta\zeta]}\delta\dot{\zeta}\rangle=-\varepsilon\label{KHM_Intertial}.
\end{equation}
Finally for an isotropic system, it can be shown the following \textit{$1$-law} for the third order structure function 
\begin{equation}
S(\ell)\equiv \langle {\bm J}_{[\delta\chi,\delta\zeta]}\delta\dot{\zeta}\rangle\cdot \hat{{\bm\ell}}=-\varepsilon\,  \ell, \label{EQ:OneLaw}
\end{equation}
where $\hat{{\bm\ell}}$ is the unitary vector along ${\bm\ell}$. {Notice that} $S(\ell)$ does not depend on any physical parameter other than the energy flux $\varepsilon$.
Note that, although $S(\ell)$ depends explicitly only on three fields ($\chi$, $\zeta$ and $\dot{\zeta}$), the Airy function $\chi$ is geometrically related to the deformation $\zeta$ by the Eq.\eqref{foppl1} ({and adequate boundary conditions). Hence}, $S(\ell)$ is thus related to a fourth order moment of the dynamical variables.

The implications of \eqref{KHM_Intertial}-\eqref{EQ:OneLaw} and the hypothesis leading to them, are important for WT and closely related to fundamental issues of HDT. We will come back to this point after validating the \textit{$1$-law} numerically.


We present now numerical simulations of equations \eqref{foppl0} and \eqref{foppl1}, that in their dimensionless form read
\begin{eqnarray} 
\frac{\partial^2 \zeta}{\partial t^2} &=& - \frac{1}{4}\Delta^2\zeta +\{\zeta,\chi\} +\mathcal{F}_0 -\nu_0(-\Delta)^{\frac{\alpha}{2}}\dot{\zeta}
\label{foppl0DimLess}\\
\Delta^2\chi &=&- \frac{1}{2}\{\zeta,\zeta\},
\label{foppl1DimLess}
\end{eqnarray}
where $\nu_0$ and $\mathcal{F}_0$ are the rescaled damping coefficient and rescaled external forcing respectively.
We supply the system with periodic boundary conditions in a square domain of size $2\pi$. The dissipative term $\nu_0(-\Delta)^{\frac{\alpha}{2}}\dot{\zeta}$ and the large-scale force $\mathcal{F}_0$ are defined in Fourier space. The forcing is white-noise in time of variance $f_0^2$ and its Fourier modes are non-zero only for wave-vectors $|{\bf k}|\le k_f$. Numerical simulations are performed using a standard pseudo-spectral code. De-aliasing is made by using the standard $2/3$-rule \cite{gottlieb1977numerical}, that is applied after computing each quadratic term. The largest wavenumber is $k_{\rm max}=N/3$, where $N$ is the resolution. In numerics we set $\alpha=6$, {$k_f=4$} and use different resolutions. All the runs of this Letter are in a statistically stationary state. The list of runs is presented in Table \ref{Table:RUNS}. {The table also displays the ratio of stretching and bending energies in the inertial range, as measure of the strength of the non-linear terms.}
\begin{table}[h]
\begin{tabular}{| l || c  | c  | c  | c  | c |}
  \hline
 	Run & 1 & 2 & 3 & 4 \\
	  \hline
	Resolution & $512^2$ & $512^2$ & $512^2$ & $1024^2$\\
	$f_0$ 			    & $14$ & $100$& $316$ & $100$\\
	$\nu_0\times 10^{-13}$ & $2.44$ & $2.44$& $2.44$ 		& $0.04$ \\
  	$E_{\rm stret}^{\rm (INE)}/E_{\rm ben}^{\rm (INE)}$ & $0.08$ & $0.25$& $0.41$ & $0.3$ \\
  \hline
 \end{tabular}
 \caption{List of runs and parameters.\label{Table:RUNS}. $E_{\rm X}^{\rm (INE)}$ is computed summing up the respective spectra within a range in the inertial zone $k\in(8,30)$.}
  \end{table}

{To verify the $1$-law we first need to determine precisely the energy flux. In  WT, due to the fact that energy is not quadratic, the fluxes can not be easily computed in Fourier space and they are typically estimated based on the injected and dissipated power \cite{humbert2013wave,Miquel2014,Deike2014}. Such methods are only approximated and useless for transient states. An exception is the determination of the energy budget scale by scale calculated in \cite{YokoyamaB} showing a  clear constant energy flux along the inertial range. We introduce an equivalent and simpler method to determine the energy flux.   For a thin elastic plate, as each term in the energy is positive (see Eqs.\eqref{Hamiltonian0}-\eqref{Energy}), the energy fluxes can be straightforwardly defined in Fourier space. Such formulas are quite analogous to those used in HDT \cite{frisch1996turbulence}. We show now how the different fluxes can be computed in the case of the FvK equations. The generalisation to other wave systems is straightforward.  }

The cross spectrum $E_{f\,g}(k)$ of two fields $f$ and $g$ is defined in terms of their Fourier transforms $\hat{f}$ and $\hat{g}$ as 
$E_{f,g}(k)=\sum_{|{\bf p}|=k} \hat{f}_{\bf p}\hat{g}_{\bf -p}$. Note that by Parseval theorem we have $\int f({\bf x})g({\bf x})\mathrm{d}{\bf x}=(2\pi)^2\,\sum_k E_{fg}(k)$. Using this definition the amplitude spectrum is $E_{\zeta,\zeta}(k)$. It relates with the standard definition of WT as $E_{\zeta,\zeta}(k)=2\pi k \langle |\hat{\zeta}_{\bf k}|^2\rangle$. The kinetic, bending and stretching energy spectra are defined as
$E_{\rm kin}(k)=\frac{1}{2}E_{\dot{\zeta}, \dot{\zeta}}(k)$, $E_{\rm ben}(k)=\frac{1}{8}E_{\Delta \zeta,\Delta \zeta}(k)$ and $E_{\rm stret}(k)=\frac{1}{2}E_{\Delta \chi,\Delta\chi}(k)$ respectively. 

Once the different energy spectra are defined, the fluxes can be determined by simple {variation of the fields} (see for instance \cite{frisch1996turbulence}). By making a standard scale-by-scale energy budget, the energy fluxes are expressed as
\begin{equation}
\varepsilon_{\rm X}(k)=-\sum_{p=0}^k \left.\frac{\partial E_{\rm X}(p)}{\partial t}\right|_{\rm NL}
\end{equation}
 where the label X stands for \textit{kin}, \textit{ben} and \textit{stret} and NL for the time variation of the fields coming only from the Hamiltonian terms (excluding forcing and dissipation). The latter is not a total time derivative when forcing or dissipation are present, therefore they do not necessarily vanish in a steady state. {The energy fluxes are obtained by direct calculation and they read:}
 \begin{eqnarray*}
&&\varepsilon(k)=\varepsilon_{\rm kin}(k)+\varepsilon_{\rm ben}(k)+\varepsilon_{\rm stret}(k),\\
&&\varepsilon_{\rm kin}(k)=-\sum_{p=0}^k E_{\dot{\zeta} , \{\zeta,\chi\}}(p)+\frac{1}{4}\sum_{p=0}^k E_{\Delta\zeta ,\Delta\dot{\zeta}}(p),\\
&&\varepsilon_{\rm ben}(k)=-\frac{1}{4}\sum_{p=0}^k E_{\Delta\zeta ,\Delta\dot{\zeta}}(p),\,\,\varepsilon_{\rm stret}(k)=\sum_{p=0}^k E_{ \chi,\{\zeta,\dot{\zeta}\}}(p).\hspace{.6cm}		
\end{eqnarray*}
{For instance, we have that $\varepsilon_{\rm stret}(k)=\sum_{p=0}^k E_{ \chi,\{\zeta,\dot{\zeta}\}}(p)$, and as $E_{\Delta \chi,\Delta \dot{\chi}}(p)=E_{\chi,\Delta^2 \dot{\chi}}(p)=-E_{ \chi,\{\zeta,\dot{\zeta}\}}(p)$, the above formula follows.}
Note that because of the energy conservation by the Hamiltonian dynamics we have $\lim_{k\to\infty}\varepsilon(k)=0$. In numerics, if (and only if) the code is correctly de-aliased, we have $\varepsilon(k_{\rm max})=0$.

\begin{figure}[h]
\centering
\includegraphics[width=0.49\textwidth]{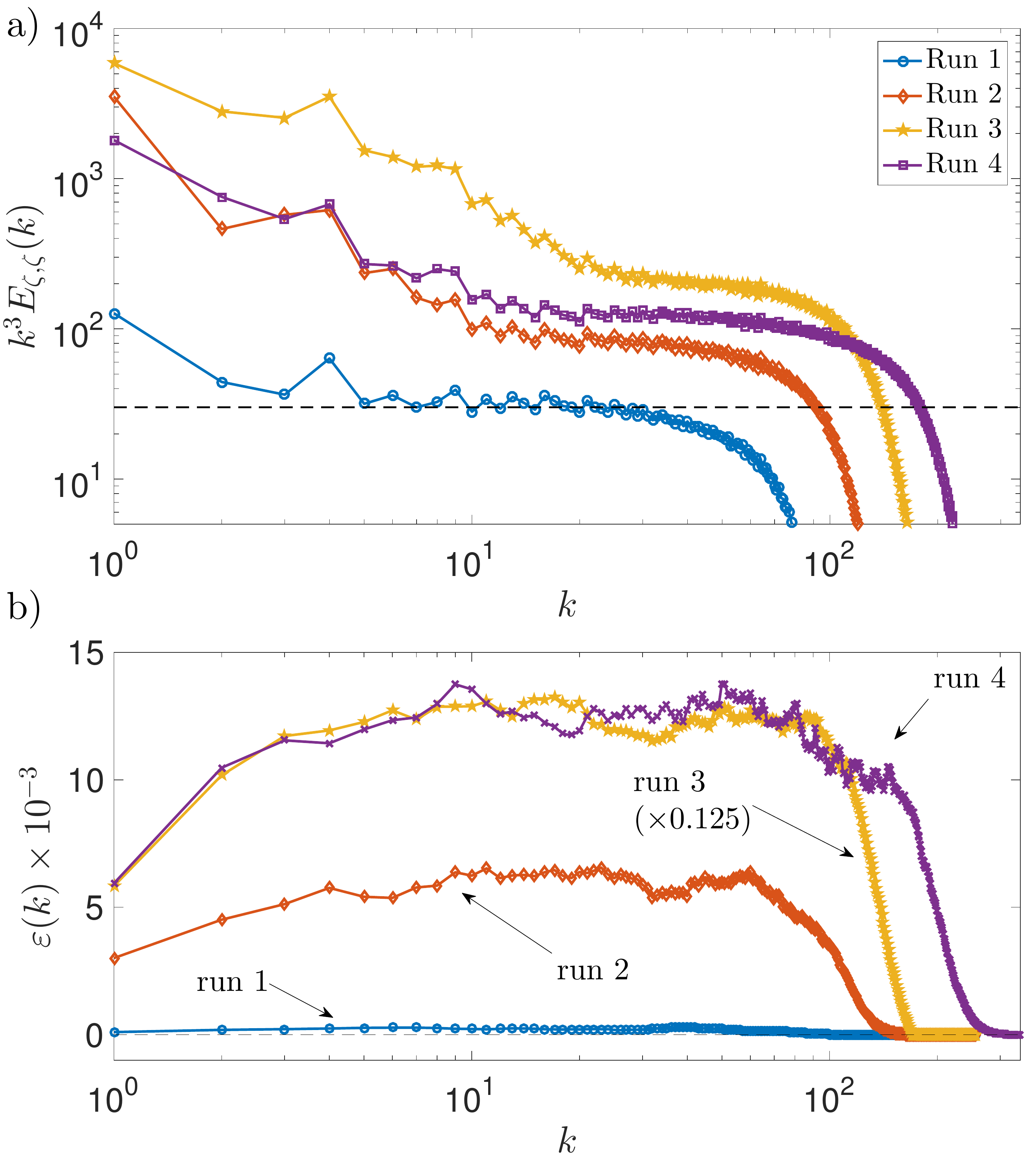}
\caption{(Color online) a) Different amplitude spectrum. The dashed line displayed the weak WT theoretical prediction. Runs 1 and 4 present a good agreement with theoretical predictions (data not shown for the sake of clarity in the figure) b) Different fluxes for all runs. The spectra and the fluxes have been time-avaraged in the statistical stationary state. The fluxes have been normalized by the mean value of the flux $\bar{\varepsilon}$ measured in the inertial range of each run.}
\label{Fig:spectral}
\end{figure}

We present now our numerical results. Figure \ref{Fig:spectral}.a displays the amplitude spectra $E_{\zeta,\zeta}(k)$ {compensated by $k^3$} for different runs. 
The dashed line indicates the scaling $k^3E_{\zeta,\zeta}(k)\sim k^{0}$ predicted by the weak WT theory \cite{placa2006,during2017wave}.
Theoretical prediction agrees well {for run 1 that corresponds to the one in the weaker non-linear regime, whereas the others runs display a steeper spectra}, indicating the possibility of strong wave turbulence as in \cite{ChibbaroJosserandPlateIntermittency}. In order to verify if the scaling observed in Fig.\ref{Fig:spectral}.a corresponds to a cascade process with a constant flux in the inertial range, the (time-averaged) fluxes are presented in Fig.\ref{Fig:spectral}.b for all runs. They are all flat in the inertial range. 

\begin{figure}[h]
\centering
\includegraphics[width=0.49\textwidth]{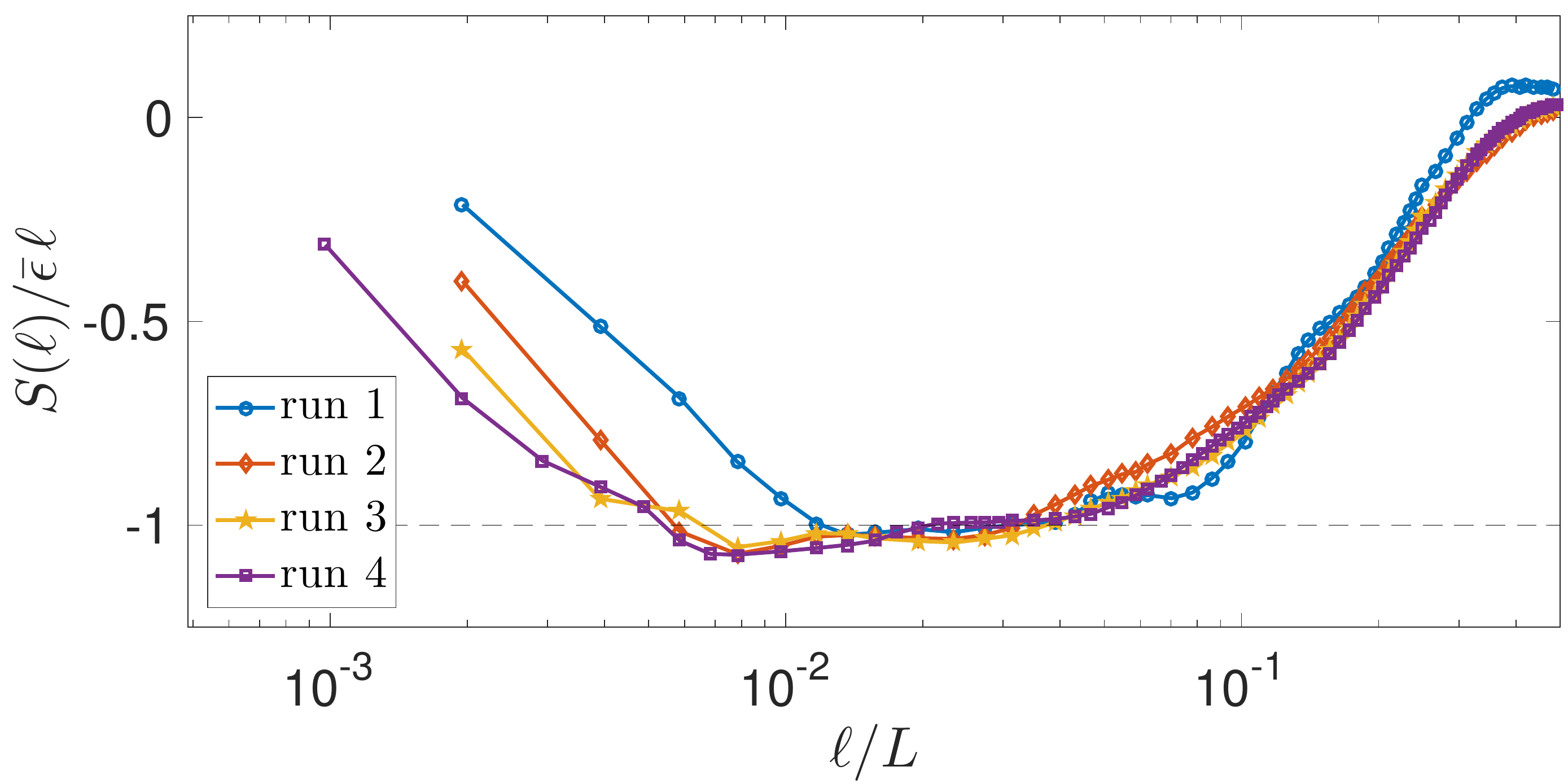}
\caption{(Color online) Normalized structure function $S(\ell)$ defined in Eq. \eqref{SFunction} as a function of the scale $\ell$. $\bar{\varepsilon}$ is measured for each run measuring the flux in the inertial range. $L$ is the size of the domain. The theoretical prediction \eqref{EQ:OneLaw} is represented by the horizontal dashed line.}
\label{Fig:OneLaw}
\end{figure}

We proceed now to verify the main result of this Letter, namely the \textit{$1$-law} in Eq.\eqref{EQ:OneLaw}. For each run we mesure the value $\bar{\varepsilon}$ directly averaging the energy flux in the inertial range. The structure functions $S(\ell)$ normalised by $\bar{\varepsilon} \ell$ are displayed in Fig.\ref{Fig:OneLaw}.
The theoretical prediction \eqref{EQ:OneLaw} is displayed in excellent agreement by the black dashed line.

Besides the standard assumptions of homogeneity and isotropy, the derivation of the K\'arm\'an-Howarth-Monin relation \eqref{KHM_Intertial}-\eqref{EQ:OneLaw}  assumed that the rate of energy dissipation remains finite when the scale separation between injection and dissipation of energy tends to infinity (for instance making $\nu_0\to0$ in \eqref{foppl0DimLess}). In the context of 3D incompressible HDT driven by the Navier-Stokes equations, this fundamental property is known as the \textit{dissipative anomaly} \cite{frisch1996turbulence}. It is related to the Onsager's conjecture that the remanent dissipation in the limit of infinite Reynolds number can be associated with singular (weak) solutions of the Euler equation that do not conserve energy \cite{eyink2008dissipative}. To our knowledge, such fundamental questions have not been yet addressed in the context of the F\"oppl-von K\'arm\'an equations. It would be of great interest to investigate (theoretically, numerically and experimentally) if such anomaly exists in WT of thin elastic plate and other related systems.

We would like to emphasize that the 1-\textit{law} in Eq.\eqref{EQ:OneLaw} is valid for both, weakly and strongly interacting waves. It is interesting to notice that a {naive} scaling argument {would} suggest a contradiction with weak WT theory. From weak WT theory the amplitudes $\zeta$ are expected to scale with the energy flux as $\varepsilon^{1/6}$, what would lead to a structure function in \eqref{EQ:OneLaw} scaling as $\varepsilon^{2/3}$,  in  contradiction with the $1$-\textit{law}.  A way to conciliate this contradiction is that an exact cancelation at the leading order take place, and high order terms of the weak WT theory are needed to be taken into account. Such calculation have not been yet performed and is out of the scope of this Letter.
Finally, in the limit of $\l\to0$, where the weak WT theory breaks down, waves are absent and there is no a small parameter. We believe that the analogy between HDT and strong thin plate WT is worth to be developed further. In this limit  it is expected that d-cones and ridges appear \cite{Miquel2013}. Their effects on the energy transfers and the $1$-\textit{law} are unclear.  In this spirit, whether the limits of time going to infinity, and dissipation and thickness of the plate going to zero commute or not, it remains a fundamental and open question. The K\'arm\'an-Howarth-Monin relation \eqref{KHM} and the $1$-\textit{law} \eqref{EQ:OneLaw} derived in this Letter should represent a theoretical benchmark for future studies on elastic turbulence and intermittency.

\begin{acknowledgments}  
The authors were supported by the Chilean-French scientific exchange program ECOS-Sud/CONICYT number C14E04. The authors also acknowledge partial support from  FONDECYT grant No. 1150463.

\end{acknowledgments}

\bibliographystyle{apsrev4-1}

%

\end{document}